\documentclass[aps,prb,twocolumn,groupedaddress, showpacs,floatfix]{revtex4}

\usepackage{graphicx}

\begin{document}

\title{Critical dynamics of superconducting BSCCO films}

\author{K. D. Osborn}
\altaffiliation[Present address: ]{National Institute of Standards
and Technology, Boulder, CO, 80305}
\author{D. J. Van Harlingen}
\affiliation{Department of Physics and
Materials Research Laboratory}

\author{Vivek Aji, Nigel Goldenfeld, S. Oh
and J. N. Eckstein}
\affiliation{Department of Physics\\University
of Illinois at Urbana-Champaign
\\1110 West Green Street\\
Urbana, IL, 61801-3080\\}

\date{\today}
\bigskip

\begin{abstract}
We report on a systematic investigation of the critical
fluctuations in the complex conductivity of epitaxially-grown
Bi$_{2}$Sr$_{2}$CaCu$_{2}$O$_{8+\delta}$ films for ${T \lesssim
T_c}$ using a two-coil inductive technique at zero applied field.
We observe the static 3D-XY critical exponent in the superfluid
density near $T_{c}$. Linear scaling analysis close to the
critical temperature yields a dynamic critical exponent of $z
\approx 2.0$ for small drive currents, but non-linear effects are
seen to be important. At $T_{c}$, a non-linear scaling analysis
also yields a 3D dynamic exponent of $z = 2.0\pm 0.1$.
\end{abstract}


\maketitle

\section{Introduction}

Direct observation of the scaling properties of the
superconducting transition has at last become feasible due to the
discovery of the cuprate superconductors.  Microwave measurements
in ultrapure single crystal YBa$_2$Cu$_3$O$_{7-x}$ (YBCO) have
provided strong evidence that for accessible temperature ranges,
the effective static universality class is the three dimensional
XY model (3D-XY)\cite{kamal}, where the penetration depth
displayed scaling over three decades in reduced temperature.
Measurements on thin films have yielded varying results: microwave
measurements \cite{booth} report $\nu = 1.2$, contactless ac
conductivity measurements \cite{naki} yield $\nu = 1.7$ at
frequencies up to $2$ GHz and dc conductivity in finite magnetic
fields \cite{robe} give $\nu = 0.9-1.0$. While none of these agree
with the 3D-XY scaling behavior, it is widely believed that the
results on pure single crystals are indicative of the intrinsic
critical fluctuations of the superconducting transition.

Transport measurements have probed the dynamics of the critical
fluctuations, which are characterized by the value of the dynamic
critical exponent $z$ describing how the relaxation time $\tau$
scales with the correlation length $\xi$: $\tau \sim \xi^z$.
However for YBCO, experiments have not yielded a consistent
picture. For example, longitudinal dc-resistivity measurements
\cite{JTK} yield $z = 1.5 \pm 0.1$ in magnetic fields up to $6$T,
while microwave conductivity measurements \cite{booth} are
consistent with $z = 2.3-3.0$. Still larger values for the dynamic
exponent ranging from $z = 5.6$ to $z = 8.3$, reminiscent of
glassy dynamics, have also been seen in resistivity measurements
\cite{naki,robe} in thin films. These large values of $z$ strongly
suggest that disorder plays a dominant role in these thin films,
and this may explain why their observed static critical behavior
differs from that of the best single crystals.

Critical fluctuation results on Bi$_{2}$Sr$_{2}$CaCu$_{2}$O$_{8+\delta}$
(BSCCO) are even more dissimilar, partially due to highly anisotropic
fluctuations. The complex conductivity of underdoped BSCCO films measured at
terahertz frequencies was found to agree with a Kosterlitz-Thouless-Berezinskii
(KTB) model with diffusive dynamics.\cite{corson,minnhagen} In previous work on
BSCCO crystals a non-linear transport study found 2D critical behavior with an
anomalous dynamic exponent of $z \approx 5.6$ and no crossover to 3D
behavior.\cite{pierson} The 3D universal phase angle of the complex
conductivity deduced from the magnetic susceptibility of BSCCO epitaxial films
and crystals has even yielded larger values of $z$.\cite{kotzler} Only dc
fluctuation conductivity measurements in BSCCO \cite{rapp} find relaxational
dynamics, $z \approx 2.0$, {\it assuming\/} a 3D-XY static exponent, $\nu
\approx 2/3$, in agreement with recent theoretical work that predicts that $z
\approx 2.0$ in 3D.\cite{aji} These results indicate the lack of consensus on
the true nature of the dynamic fluctuations representative of the
superconducting transition, which is the primary motivation of the present
investigation. While a large value of the dynamic exponent $(z>2)$ can be
understood as disorder effects, only the presence of a previously undetected
collective mode coupled to the superfluid density would lead to $z \approx 1.5$
in 3D, just as occurs in superfluid He$^4$.\cite{HH}

In this paper we provide results on the critical fluctuations in
high quality epitaxially-grown BSCCO films. We have measured the
ab-plane complex conductivity with a two-coil inductive technique
at zero applied magnetic field ($H=0$), described in Section II.
Our measurements, presented in section III, indicate that the
critical fluctuations in BSCCO films are consistent with the 3D-XY
critical fluctuation model, rather than a KTB transition within
the layers. This contrasts with earlier studies using this
technique on YBCO, which have reported KTB critical \cite{fio} and
mean-field \cite{lin} fluctuations. In section IV, we estimate the
temperature intervals in which possible crossovers to 2D
fluctuations can occur, due either to decoupling of the bilayers
or to the thin film behaving as a 2D system with a large $c$-axis
correlation length.  We find that our measurements should indeed
be well-described by anisotropic 3D-XY critical fluctuations.   In
section V, we provide a crossover analysis of the approach to the
critical point, and examine the approach to a universal phase
angle. In particular, we predict the qualitative forms of the
variations with temperature and frequency.  These are in agreement
with our measurements, but contrast earlier results which report
very different values of the static and dynamic critical exponents
than those reported here. Analysis of the phase angle of the film
response using linear response theory in the critical region,
which yields a dynamic exponent of $z = 2.0$, is presented in
section VI. However, the breakdown of linear response theory
becomes important near the transition temperature $T_{c}$, so that
the analysis is complicated by the need to extract the superfluid
density from the measured mutual inductance by an approximate
inversion technique. Our analysis shows that the determination of
$z$ purely from linear response theory is contaminated by
non-linear effects, and at probe currents that are too large the
effective value of $z$ inferred can be as low as 1.5. To
circumvent this problem, we derive in section VII a non-linear
scaling law obeyed by the raw measured mutual inductance at
$T_{c}$ (i.e. without the need for a data inversion) and extract
the dynamic exponent directly from the mutual inductance. In
agreement with the linear response analysis, we again obtain $z
\approx 2.0$. Section VIII summarizes our conclusions.

\section{Samples and Experimental Setup}

We have measured several high quality oxygen-doped BSCCO films,
grown by molecular-beam-epitaxy on SrTiO$_{3}$ substrates and
analyzed \textit{in-situ} by RHEED \cite{ECK}. The BSCCO films are
nominally oriented along the c-axis, with unit cell thickness of
$d = 15.4$ {\AA}.  AFM images of similar BSCCO films reveal an
\textit{rms} roughness of 5 Angstroms primarily due to single unit
cell step height variation over the surface.  The \textit{rms}
roughness of the substrates are approximately 1.5 Angstroms.

The inductively-measured $T_{c}$ from several films of different
dopings was fit to the empirical formula $T_{c}=T_{c, max}
(1-82.6(p-0.16))^2$, to determine the hole doping per Cu,
p.\cite{presland}  The two branches of the formula are
distinguished by the relative low temperature superfluid density,
which is known to increase monotonically through optimal doping
(p=0.16). Films A, B, and C have $n = 21, 40$ and $60$ unit cells
respectively and are grown on $10$ and $14$ mm square substrates.
Film A is overdoped at $p\simeq0.19$, while film B and C are
slightly overdoped and underdoped, respectively. We estimate that
a doping gradient in the samples introduces a experimental
temperature broadening of 0.15 K near $T_c$.

To obtain the ab-plane complex conductivity, we employ a two-coil
inductive technique.  The mutual inductance of two coaxial coils
is measured with a film placed in between and normal to the axis
of the coils.  An ac current is applied through the drive coil; a
second coil, attached to a large input impedance lock-in
amplifier, measures the fields produced by the drive coil and the
screening currents in the film. The coils have $135$ turns and an
average radius of $1.5$ mm. The drive coil is placed $0.35$ mm
above the film so that at a current of $I = 50 \mu$A \textit{rms},
the applied ac magnetic field is $< 0.1$ Gauss normal to the film.
Measurements are performed in a He$^4$ continuous-flow dewar,
which allows the temperature of the film to be controlled without
significant heating from the drive coil. The sample temperature is
monitored by a Si diode attached to the back side of the
substrate.

The complex conductivity $\sigma  = \sigma _1  - i\sigma _2 $ of each film is
calculated from the mutual inductance using the exact analytical expression for
an infinite diameter film \cite{clem} and a numerical inversion technique
similar to that of Fiory \textit{et al.} \cite{fiory}. The method of Turneaure
\textit{et al.} \cite{turneaure} is employed to correct for film diameter
effects prior to calculating the in-plane complex conductivity. The measured
penetration depth $\lambda _{ab}  = (\mu _0 \omega \sigma _2 )^{ - 1/2} $, is
independent of the measurement frequency from ${f=10-100}$ kHz below the
transition temperature, where the dissipation is small ($\sigma _1 \ll \sigma
_2) $.  To analyze the critical fluctuations, we study the complex superfluid
density per CuO bilayer,
$\rho = i\sigma\omega d\Phi_{0}^{2}/(4\pi^{2}k_{B})$,
where $\Phi _0 $ is the superconducting flux quantum.\cite{FFH}
The real part of this quantity, $Re[\rho]$ has units of
temperature and sets the energy scale for critical fluctuations.
In Fig. 1(a), $Re[\rho]$ obtained at $f = 80$ kHz and $I = 40
\mu$A is plotted for films A through C.  At $5$ K, films A, B, and
C, have penetration depths of $235$, $265$, and $285$ nm,
respectively. Note that film A, which is overdoped, has a smaller
low temperature penetration depth than film B and C. All films
exhibit a $T^2$ component to the superfluid density at low
temperatures, however the zero-temperature intercept of the films
near optimal doping is approximately zero, indicating the
crystalline films have very few impurities.

\begin{figure}[h!]
\includegraphics[width=8cm, bb=10 84 570 780]{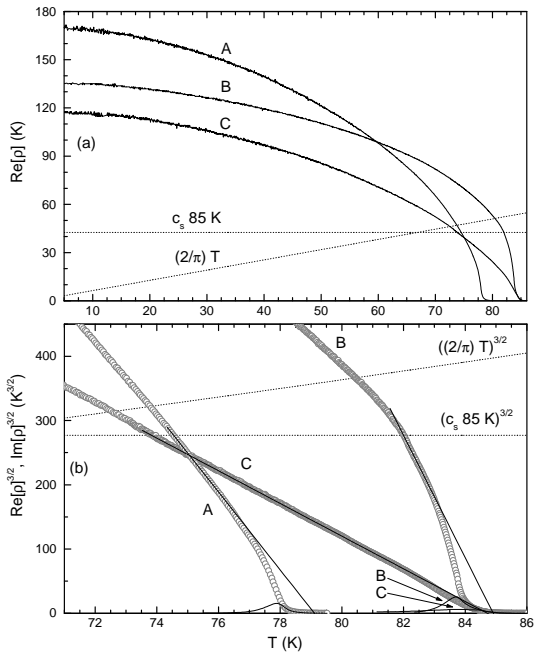}
\caption{\label{fig1}$Re[\rho]$ (panel a) and $Re[\rho]^{3/2}$ (panel b), as a
function of temperature for films A through C. $\rho=(c_{s}85 $K) is the
predicted crossover to 3D-XY critical behavior and $\rho = (2/\pi)T$ is the KTB
critical superfluid density for an isolated bilayer.}
\end{figure}

\section{Static critical behavior}

In Figs. 1(a) and 1(b) a power of $\rho=Re[\rho]$ and $ 2T/\pi$ is
shown to locate where we would expect the KTB transition for a
noninteracting bilayer, $\rho = 2T/\pi$.\cite{bea}  We observe no
such drop in the superfluid density at this temperature in our
films. The films instead show that coupling among the bilayers
induces an anisotropic 3D transition, precluding a KTB transition
in the bilayers.  At the transition temperature of an isolated
bilayer $T=\rho\pi/2$, however, the measured superfluid density
has been renormalized from the mean field value by quasi-2D
fluctuations.

The mean field behavior, expected at low temperatures, crosses
over to the 3D-XY behavior as the critical temperature is
approached from below. In layered systems, the 3D-XY behavior is
exhibited once the c-axis correlation length exceeds the
interlayer spacing, $d$; i.e. when $\rho(T) = c_{s}T_{c}$, where
$c_{s} \approx 0.5$ and $\rho=Re[\rho]$.\cite{kamal} In Fig. 1(b),
the data from Fig. 1(a) are plotted as $Re[\rho]^{3/2}$ near the
critical regime. Also shown in Figs. 1(a) and 1(b) is the
corresponding power of $c_{s}T_{c}$ to locate the onset of 3D-XY
behavior. The static 3D-XY model gives $\rho \sim \xi^{-1} \sim
|T-T_{c}|^{\nu}$, where $\nu \approx 2/3$. We observe that
$Re[\rho]^{3/2}$ varies linearly in temperature for $\rho(T)
\lesssim c_{s}T_{c}$ indicating a 3D-XY, rather than mean-field
($\rho \sim |T-T_{c}|$), behavior. At our measurement current ($40
\mu$A), we observed an absence of frequency dependence below the
dissipation peak over the temperature range of the fit. In film C,
the linear regime extends as far as 7 K below $82$ K. The linear
fits extrapolate to $T_{c} = 79.1$ K , $84.9$ K and $84.7$ K for
films A, B and C respectively.

Although our static scaling analysis is carried out at
temperatures and at a drive current for which there is no
discernible frequency dependence, at higher temperatures we
observe variations with both drive current and frequency. This is
apparent in Fig. 2, for which larger currents are used on film B
with 40 layers. Film C, with 60 layers, exhibited a weaker
dependence over the same frequencies and currents. Since the
superfluid density depends on frequency and current at higher
temperatures, the superfluid density scales as $\rho \sim
\xi^{2-D}f(\omega\xi^{z},E\xi^{1+z})$, where E is the amplitude of
the electric field.  Only in the critical regime where the data
are independent of both frequency and electric field can one
expect to see the linear response scaling behavior and extract the
exponent $\nu$. At higher temperatures where we observe non-linear
behavior as well as frequency dependence, the data are still
consistent with the 3D-XY fluctuations, and we can extract $z$. In
section VI we analyze the phase angle in terms of linear response
theory, but a nonlinear scaling analysis is required for
completeness and is given in section VII.

\begin{figure}[h!]
\includegraphics[width=8cm, bb=60 60 520 700]{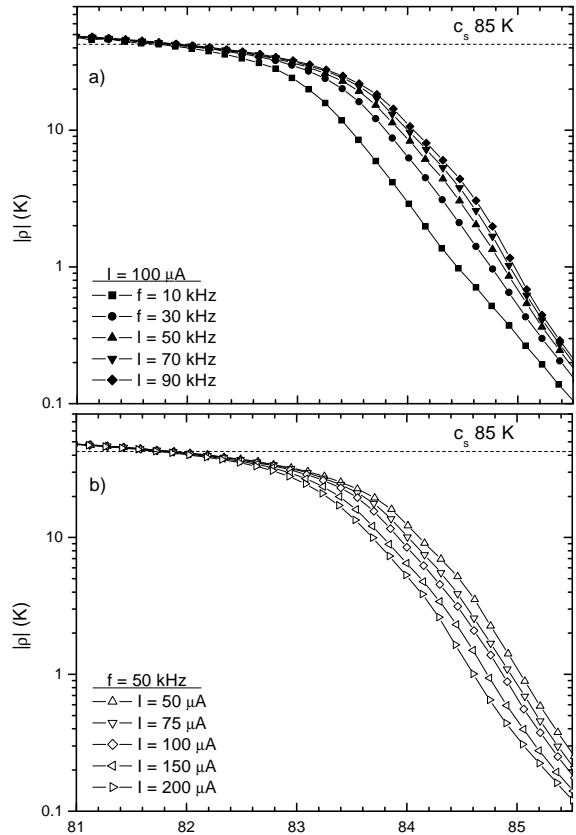}
\caption{\label{fig2}$|\rho|$ in film B taken at a series of frequencies (panel
a) and drive currents (panel b).}
\end{figure}

\section{Dimensionality of the static fluctuations}

When the superfluid density is a fraction of the 3D crossover
($\rho(T) < (1/2) c_{s}T_{c}$), we observe a drop in the
superfluid density from the static scaling value. Although this
could be interpreted as a crossover to 2D behavior, we do not
believe this to be the case in our films since the drop is
observed to be current as well as frequency dependent (see Fig.
2). Notice that the superfluid density does drop more rapidly than
expected from the 3D-XY model as the frequency is decreased for a
given drive current, but does so more slowly as the current is
decreased at fixed frequency. While the former might lead one to
conclude that the anisotropic 3D-XY behavior is no longer valid in
the static limit, the latter would imply the opposite. More
accurate measurements are needed at low frequency and currents to
establish unambiguously the true nature of the static limit.
Nevertheless the trend with decreasing current, coupled with the
phase angle dependence discussed in the next section, is
consistent with 3D-XY behavior. Another putative crossover to 2D
behavior is expected when $\xi_{c}$ becomes as large as the film
thickness, $h$. The 3D-XY scaling implies that this occurs at a
temperature, $T^{*}$, which is $0.05$ K, $0.01$ K and $0.02$ K
below the transition temperature for films A, B and C
respectively. This implies that the region of pure 2D fluctuations
is too small to be resolved in our data and therefore we expect 3D
fluctuations to dominate.

\begin{figure}[h!]
\includegraphics[width=8cm, bb=84 89 506 725]{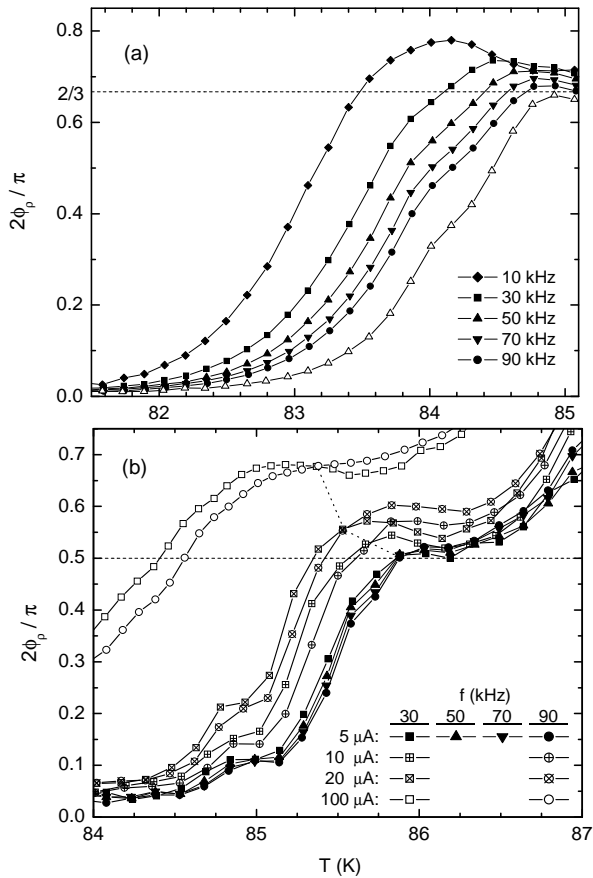}
\caption{Normalized phase angle, $2\phi_{\rho}/\pi$, of the superfluid density
extracted from linear response theory.  Panel a: $2\phi_{\rho}/\pi$ in film B
taken at $I = 100 \mu$A (filled symbols) at a series of frequencies and $I = 50
\mu$A for $ f = 50 $kHz (unfilled triangles).  Panel b: $2\phi_{\rho}/\pi$ in
film C for a given table of frequencies and drive currents. }
\end{figure}

\section{Crossover phenomena in the linear phase angle near $T_c$}

Since the experiments are performed at finite frequency, it is
important to consider whether or not the correct asymptotic behavior is
being probed and what conclusions can be drawn from the observed
behavior. We begin with the
superfluid density, within linear response, written as

\begin{equation}
\rho = (t^{-\nu})^{2-D} f_1(\omega t^{-\nu z})
\end{equation}

\noindent where $t = \left| T - T_{c} \right| /T_{c}$ and $f_1$ is a scaling
function. In the limit of $t \rightarrow 0$ ($T \rightarrow T_{c}$), the
$t$-dependence must cancel out of the expression for $\rho$, which implies that
the asymptotic behavior of the scaling function $ f_1(x) \sim x^{\theta}$ as
$x\rightarrow \infty$ where $\theta = (2-D)/z$. For $ t \neq 0$, we can write

\begin{equation}
\rho =  (i\omega)^{(D-2)/z} f_{2} (t \omega^{-1/\nu z})
\end{equation}

\noindent where the scaling function $f_{2} (y) = $ const as $y
\rightarrow 0$. We now ask what behavior we should observe as
$\omega \rightarrow 0$ for $t \neq 0$. For $\omega \rightarrow 0$
at $t = 0$, we should see the critical point behavior. For
$t/\omega^{1/\nu z} \ll 1$ we are probing the $ y \sim 0$ regime
of the scaling function $f_{2} (y)$ and expect to see asymptotic
scaling behavior. If $t/\omega^{1/\nu z} \gg 1$ we are probing the
$y \rightarrow \infty$ regime of the argument of $f_2$ and do not
expect to see the scaling behavior. Thus at a finite $t_{0}$, for
$\omega \gg t_{0}^{4/3}$ we would observe critical behavior while
for $\omega \ll t_{0}^{4/3}$ we would not. We have used the
critical exponents $\nu =2/3$ and $z =2$ for specificity, but the
argument does not rely on these precise values.  Note that, as is
generally the case with crossover, the behavior of the superfluid
density away from the critical point at finite frequencies is not
universal except in the large frequency limit - a result that is
sometimes found to be surprising, but which follows from a generic
crossover analysis of relevant variables. \cite{goldenfeldbook}
This predicted low frequency behavior is consistent with our data
where we observe the critical phase angle in Fig. 3(b) and yet we
do not observe scaling below $T_c$. More importantly, this is also
consistent with the non-linear scaling observed at $T_c$ in
Section VII. Away from $T_c$ an asymptotic scaling regime is
demonstrated in the microwave measurements of Booth {\it et
al}.\cite{booth}

\section{Dynamic critical behavior and the breakdown of linear response theory}

Next we study the phase of $\rho$, $\phi_{\rho}$, close to the
transition temperature. The frequency dependence of
$2\phi_{\rho}/\pi$ for film B measured at $I = $100 $\mu A$ is
shown in Fig. 3(a).  Within linear response theory, the phase
angle is independent of frequency at the critical temperature and
is given by $\phi_{\rho}(T_{c}) = \pi/(2z)$ in three dimensions,
where $z$ is the dynamic critical exponent.\cite{DOR} In Fig. 3(a)
we notice that the curves for different frequencies seem to
approach a phase angle consistent with a dynamic exponent of $z
\approx 1.5$ and $T_{c} = 84.9 $ K. When repeated at a smaller
value of the current, $I = 50 \mu A$, $\phi_{\rho}(T_{c})$ is
smaller, as shown. Since the constant phase angle is a result from
linear response theory, the estimate of the dynamic exponent from
the data is expected to be more accurate at smaller currents.

This is indeed seen in the current dependence of the phase angle at different
frequencies for film C. In Fig. 3(b), $2\phi_{\rho}/\pi$ is shown for a set of
frequencies and currents. As the current is lowered, the frequency independent
phase angle shifts to lower values. At $20 \mu$A, $10 \mu$A and $5 \mu$A, this
phase angle is $0.57\pi /2$, $0.54\pi /2$ and $0.51\pi /2$ respectively. This
suggests that $z \simeq 2.0 $ in film C and the result of $z \approx 1.5$ from
film B is an artifact of the large current used in the measurement. With this
analysis, the thickest film (C) exhibits the best estimate of the critical
phase angle, the film with intermediate thickness (B) provides evidence for
similar behavior, but the thinnest film did not show a critical phase angle.
Due to the crossover phenomena discussed in the previous section, the behavior
of the phase angle away from $T_{c}$ is not universal at low frequencies, but
we can still infer the dynamic critical exponent from the critical phase angle
at $T_{c}$ which is indeed universal.

Both the phase angle and the superfluid density are found to be
dependent on the drive current near $T_{c}$. In the limit of
smaller currents, the frequency independent critical phase angle
does indeed yield a dynamic exponent of $z \approx 2.0$. This is a
result consistent with the presence of $3D$ fluctuations. This is
also consistent with the behavior of the superfluid density with
decreasing current at fixed frequency, where it is seen to
approach the asymptotic static scaling law. Furthermore, the
smooth evolution of the phase angle suggests that the crossover to
the 2D regime near $T^{*}$ is not resolved in our experiment.
Nevertheless, considering that nonlinear scaling and dimensional
crossover is expected close to $T_{c}$, we now proceed to analyze
the nonlinear response to get a more accurate determination of the
dynamic exponent.

\section{Nonlinear scaling at the superconducting transition}

The field equation governing the vector
potential, $\vec{A}$, is

\begin{equation}\nabla^{2}\vec{A}/\mu_{0} = -
\vec{J_{d}} - h \delta(z) \sigma \vec{E}
\end{equation}
where $\vec{J_{d}}$ is
the current density in the drive coil, $\sigma$ is the
conductivity of the superconducting film, and $\vec{E}$ is the
electric field. For the geometry of the setup,
\begin{equation}
\vec{J_{d}} =
I_{d} \delta (r - r_{c}) \delta (z - z_{c})\hat{\phi}
\end{equation}
in cylindrical coordinates, where $I_{d}$ is the drive current,
$r_{c}$ is the radius of the drive coil and $z_{c}$ is the
distance from the film to the drive coil.\cite{clem} For an
anisotropic system, in the critical regime, the superfluid
density, $\rho$, scales as,
\begin{equation}
\rho \sim i\omega\sigma \sim
\xi_{ab}^{3-D}\xi_{c}^{-1}f(i\omega\xi_{ab}^{z}, E\xi_{ab}^{1+z})
\end{equation}
where $\xi_{ab}$ and $\xi_{c}$ are the correlation lengths parallel and
perpendicular to the CuO bilayers, and $f$ is a scaling function. We
can rewrite the field equation as
\begin{eqnarray}
\nabla^{2}A = &-& \mu_{0}I_{d}\delta(r - r_{c})\delta(z - z_{c}) \\
\nonumber &-&
h\xi_{ab}^{3-D}\xi_{c}^{-1}f(\omega\xi_{ab}^{z},A\xi_{ab})A\delta(z)
\end{eqnarray}
where we have used the scaling form for the complex conductivity. Since the
scaling function is written in terms of dimensionless variables we can non
dimensionalize the equation by recasting it in terms of $A\xi_{ab},
\omega\xi_{ab}^{z}, r/h$ and $I_{d}\xi_{ab}$. The solution then takes the form,
\begin{equation}\label{nondim_a}
A i\omega\xi_{ab}^{1+z} = G(i\omega\xi_{ab}^{z},
i\omega\xi_{ab}^{1+z}I_{d}, h^{2}\xi_{ab}^{3-D}\xi_{c}^{-1}\Lambda^{-1}, r/h)
\end{equation}
{\noindent}where $G$ is a function that can be determined by solving the full
non-linear field equation and $\Lambda$ is the thermal length given by
$\Phi_{0}^{2}/(4\pi\mu _0k_{B}T)$. So far all we have done is to use the
scaling form for the superfluid density and looked for a generic solution of a
differential equation by identifying all the non-dimensional variables. For a
given geometry of the experiment the right hand side of Eq. \ref{nondim_a} is a
function of three variables. One can further simplify the expression if we
realize that there exists a regime $\left| T - T_{c}\right| < T^{*} $ where we
are governed by the limit where $ \xi_{c} = h$. This allows us to eliminate one
of the three terms as follows. Close to $T_{c}$, using the 3D-XY behavior of
the correlation length and the definition of $T^{*}$, we can rewrite
$h^{2}\xi_{ab}^{3-D}/\xi_{c}$ as,
\begin{equation}
h^{2}\xi_{ab}^{3-D}/\xi_{c} =
h\left|1+(T-T^{*})/(T^{*}-T_{c})\right|^{\nu}.
\end{equation}
For a temperature range $\left|T-T^{*}\right| \ll \left|T^{*} -
T_{c}\right|$, $h^{2}/\xi_{c} \approx h$. Thus, for temperatures
within $0.05K$, $0.02K$, and $0.01$K of $T^{*}$ for films a, b,
and c respectively, this approximation is valid, and the field in
Eq. \ref{nondim_a} is a function of only two variables for a given
geometry. Thus one can now perform the standard data collapse at
$T_{c}$ of the measured mutual inductance to obtain the dynamic
exponent. In this regime we can eliminate $\xi_{ab}$ from
Eqn.\ref{nondim_a} so that
\begin{equation}
A =
\omega^{1/z}\widetilde{G}(I_{d}\omega^{-1/z},h\Lambda^{-1},r/h).
\end{equation}
The mutual inductance, $M$, is related to the vector potential at the
pick up coil and scales as
\begin{equation}
M \sim A\omega^{-1/z} \sim \widetilde{G}(I_{d}{i\omega}^{-1/z}).
\end{equation}

Within the resolution of our data, any data collapse observed is a
3D phenomena starting to crossover to the asymptotic 2D scaling.
The reason for this is that the solution (Eq. \ref{nondim_a})
depends on a number of variables, but sufficiently close to
$T_{c}$ there exists a regime wherein the mutual inductance is
only a function of one scaling variable. If we do not observe any
data collapse we would conclude that we are never in the regime
where the approximation $h^{2}/\xi_{c} \approx h$ is valid. It is
indeed true that the elimination of the dependence of $\xi_{ab}$
in Eq. \ref{nondim_a} is possible only at $T_{c}$ and need not
hold in the entire regime where the crossover occurs. It is
tempting to interpret the experimental results in terms of a 2D
system. Our data on the other hand are unable to resolve the
crossover regime, and yet do exhibit data collapse (see Fig 4).
This suggests that the temperature at which we observe data
collapse lies in the crossover regime, but is not necessarily the
true $T_{c}$. From our analysis we conclude that the $T_{c}$ is
within $0.02$ K of the temperature at which we observe data
collapse for film B and C. We now look for data collapse of our
data measured by varying frequency, current and temperature.
Notice that the value of the exponent $\nu$ is required only to
establish the regime of validity of the approximation and is not
necessary to extract the dynamic exponent from data collapse at
$T_{c}$.

\begin{figure}[h!]
\includegraphics[width=8cm, bb=27 10 562 764]{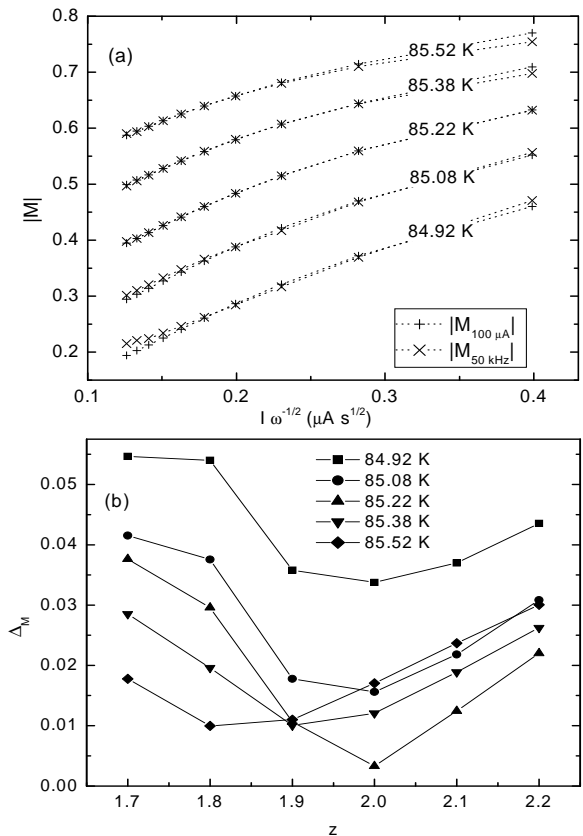}
\caption{Panel (a): $|M|$ measured at $I = 100 \mu$A and $f = 50$ kHz versus
the scaling variable $I\omega^{-1/2}$. Panel (b): Scaling error at different
temperatures versus $z$.}
\end{figure}

For the scaling analysis at $T_c$, the mutual inductance at fixed
frequency, $M_f$, was compared to the mutual inductance at fixed
current, $M_I$. For convenience, $|M(T)|$ is normalized to unity
above the $T_c$.  Values of $M_f$ ($M_I$), were measured at 12.5
$\mu$A (10 kHz) intervals for a set of fixed temperatures. Then
values of $|M_f|$ were compared with values of $|M_I|$ taken at
the same temperature by fitting the raw data of $|M_f|$ to a
polynomial and selecting points with the values of
$I\omega^{-1/z}$ equivalent to those of $|M_I|$. In Fig. 4(a)
$|M_f|$ and $|M_I|$ for film B are plotted versus $I\omega^{-1/2}$
for a series of temperatures. The best data collapse of the curves
appear at $z=2.0$ and $T=85.22$ K. The error in the scaling,
$\Delta _M $, is shown in Fig. 4(b) for different temperatures and
values of $z$.  This is obtained from $\Delta _M^2 = \sum \left(
{\left| {M_f } \right| - \left| {M_I } \right|} \right)^2 $, where
the sum runs over the 10 values of $I\omega^{-1/z}$ measured for
$|M_I|$. The lowest error value indicates $z = 2.0\pm 0.1$ and $T
= 85.2 \pm 0.1$ K. A similar analysis for the film C yields $z =
1.8 \pm 0.2$ and $T = 85.9 \pm 0.2$ K, which is also consistent
with $z = 2.0 \pm 0.1$. The transition temperature of film B is
$T_{c} = 85.2 \pm 0.1$K, while for film C it is $T_{c} = 85.9 \pm
0.2$ K. These values of $T_{c}$ are close to those obtained with
the phase angle measurements.

In Fig. 4(b), the evolution of the error for a fixed trial $z$
also suggests that the dynamic exponent observed is indeed due to
$3D$ fluctuations. Notice that as the temperature is increased for
$z=2.0$, the error decreases, approaching the best fit (i.e. data
collapse), and then increases again. Within the resolution of the
data, we never observe any change to the $2D$ regime. In other
words, the approximation that allowed us to look for the data
collapse is indeed supported by the behavior of the mutual
inductance.

\section{Conclusions}

The superfluid density in BSCCO films with different thicknesses
has been measured and shown to be consistent with the predictions
of the 3D-XY model. The dynamic scaling exponent has been obtained
by performing both a linear and non-linear scaling analysis. While
the linear scaling analysis did yield a 3D dynamic exponent of $z
\approx 2.0$, non-linear effects needed to be studied to get an
accurate determination. This is evident from the current
dependence of the critical phase angle of the superfluid density.
A 3D non-linear scaling analysis also yields a dynamic exponent of
$z = 2.0 \pm 0.1$.

\medskip

\begin{acknowledgments}
We thank H. Westfahl and D. Sheehy for many helpful discussions.
This work was supported by the U. S. Department of Energy,
Division of Materials Sciences, grant DEFG02-91ER45439 through the
Frederick Seitz Materials Research Laboratory at the University of
Illinois at Urbana-Champaign, and by the National Science
Foundation under grant number NSF-DMR-99-70690.
\end{acknowledgments}

\end{document}